\documentclass[12pt]{article}
\usepackage{amsfonts} 
\usepackage{a4}
\usepackage{epsfig}
\usepackage{latexsym}
\begin{document}

\title {Vacuum energy in the presence of a magnetic string with delta function profile}
\author{Marco Scandurra \thanks{e-mail: scandurr@itp.uni-leipzig.de} \\
  Universit\"at Leipzig, Fakult\"at f\"ur Physik und Geowissenschaften \\
  Institut f\"ur Theoretische Physik\\
  Augustusplatz 10/11, 04109 Leipzig, Germany} \maketitle

\begin{abstract}
We present a calculation of the ground state energy of  massive spinor fields and massive  scalar fields in the background of an inhomogeneous magnetic string with  potential given by a delta function. The zeta functional regularization is used and the lowest heat kernel coefficients are calculated. The rest of the analytical calculation adopts the Jost function formalism. In the numerical part of the work  the renormalized vacuum energy as a function of the radius $R$ of the string is calculated and plotted for various values of the strength of the potential. The sign of the  energy is found to change with the radius. For both scalar and spinor fields the renormalized energy shows  no logarithmic behaviour in the limit $R\rightarrow 0$, as was expected from the vanishing of the heat kernel coefficient $A_2$, which is not zero for other types of profiles.

\end{abstract}

\section{Introduction}
 
Recent developments in the technique for the calculation of the
zero point energy of  massive fields \cite{firstball} have opened interesting possibilities in the study of  soft boundaries and smooth potentials with spherical and cylindrical geometry  immersed in the vacuum of various quantum fields. The technique  uses the Jost function of the scattering problem related to the background potential under examination. The vacuum energy can be expressed as an integral containing the logarithm of a Jost function with imaginary argument:

\begin{equation}
E_{B}\ =\ -\frac{\cos \pi s}{\pi} \mu^{2s} \sum_l \int_{m_e}^\infty dk (k^2-m_e^2)^{1/2-s}
\frac{\partial}{\partial k} \ln f_l(ik)\ ,
\end{equation}
here the subscript $B$ represents the background potential, $\sum_l$ and $\int_k$ are respectively  the integration over the momentum $k$  and the summation over all other quantum numbers $l$; $m_e$ is the mass of the quantum field, $s$ is a regularization parameter and $\mu$ a mass parameter. 
The Jost function $ f_l(ik)$ is unique and easily obtainable for many types of geometries with circular, spherical or cylindrical symmetry and for various types of potential profiles, which is the great advantage of this approach. 

Representation (1) for the vacuum energy still needs a renormalization. The zeta functional representation of (1) and its heat kernel expansion are known to be a very effective tool in this context (see for instance \cite{Boston}). The basic idea to define unambiguously a renormalized zero point energy, is to impose the condition that the vacuum energy is zero when the mass of the quantum field reaches infinity (see  \cite{firstball}).  
 
The heat kernel coefficients themselves are also of great interest, they determine the asymptotic behaviour of the renormalized energy \cite{dielectric}; they are to be considered as an intrinsic local feature of the geometry and of the background under examination.
 
This renormalization scheme and the above mentioned integral representation have been used to solve some basic configurations: the vacuum energy of scalar, spinor and electromagnetic fields in the background of spherical shells with hard and smooth potentials has been  calculated \cite{firstball}, \cite{Marco}, \cite{Jou31}, \cite{dielectric}, \cite{Romeo}. The investigation has turned recently to magnetic fields  with cylindrical symmetry.  In paper \cite{master} a complete analysis of a spinor field in the background of an homogeneous magnetic flux tube of finite radius was carried out. The vacuum energy was found to be negative and it did not show a minimum for any finite value of the radius. A natural question is if inhomogeneous magnetic fields can minimize the energy and render the string stable. The question was already raised in \cite{Cangemi}. The present paper extends the investigation begun in \cite{master} to an inhomogeneous magnetic string with delta function profile. The delta function, although not a fully realistic physical model, represents a simple example of inhomogeneity, which could give an insight into the problem  of vacuum energy in magnetic backgrounds. This kind of ``semi-transparent'' was already analyzed in \cite{Marco} for a sphere. It has some features in common with a smooth potential and some with a hard boundary. In paper \cite{heat kernels} the heat kernel coefficients for a general semi-transparent boundary were calculated. The quantum mechanics of spinor fields in the presence of magnetic fluxes has been elaborated in early works \cite{Heisenberg},\cite{Weisskopf} while more recent  investigation in this direction has been motivated by the interest in the Aharonov-Bohm effect \cite{Serebryany}, \cite{Sitenko}.  Singular inhomogeneous magnetic fields were  examined in \cite{Fry} for the calculation of the fermion determinant and in \cite{Voropaev} for the investigation of the bound states of an electron, however the ground state energy was not calculated in those  works. In our paper the ground state energy will be calculated for a scalar and for a spinor field. In  the first part of this paper we will calculate the Jost functions for both fields, the heat kernel coefficients will be found and the energy will be renormalized imposing the vanishing of the vacuum fluctuations for fields of infinite mass. In the second part of the paper we will work numerically on the renormalized energy finding its asymptotic behaviour for small and for large values of the radius of the string. We will finally show some plots of the renormalized energy for various values of the strength of the potential.

\section {Scalar field in the background of a magnetic string}
\subsection{Solution of the field equation}
We quantize a scalar field $\Phi$ in the presence of a classical magnetic field whose form is that of a cylindrical shell with delta function profile. The section of the string is a circle with radius $R$. The magnetic field is given by 

\begin{equation}
\vec{B}(r)\ =\ \frac{\phi}{2\pi R}\ \delta(r-R)\ \vec{e}_z   
\end{equation}  
where $\phi$ is the magnetic flux, $r=\sqrt{x^2 +y^2}$ and $z$ is the axis along which the cylindrical shell extends to infinity.
 The quantum field  obeys the Klein- Gordon equation for the scalar electrodynamics:

\begin{equation}
(D^2+m_e^2)\Phi(x)=0
\end{equation}
where $m_e$ is the mass of the field, and

\begin{equation}
\begin{array}{lcl}
D^2 & = & \partial_t^2-\vec{\nabla}^2-2ieA^0\partial_t - 2ie\vec{A}\vec{\nabla} - e^2A^0\ ^2 + e^2\vec{A}^2\ , 
\end{array}
\end{equation}
here $e$ is the electron charge, $A_\mu$ is the vector potential of the electromagnetic field and  the convention $g_{\mu\nu}=$diag$(1,-1,-1,-1)$ is used as well as the gauge $\partial^\mu A_\mu=0$.
We want to find a solution of eq.(3) in cylindrical coordinates $z,r$, $\varphi$, then we take the relevant operators in (4) in the following form

\begin{equation}
\vec{\nabla} \rightarrow\ \left(-\cos \varphi\partial_r-\frac{\sin\varphi}{r}\partial_\varphi,\ \sin\varphi\partial_r+\frac{\cos\varphi}{r}\partial_\varphi,\ \partial_z\right) 
\end{equation}
\begin{equation}
\vec{\nabla}^2 \rightarrow\ \frac1r\partial_rr\partial_r+\frac{1}{r^2}\partial_\varphi^2+\partial_z^2\ . 
\end{equation}
The potential four vector $A^\mu$ associated with the magnetic field (2)  contains a theta-function: 

\begin{equation}
\vec{A}=\frac{\phi}{2\pi}\frac{\Theta(r-R)}{r}\vec{e}_\varphi\ , \ \ A^0=0\ ;
\end{equation}
 
where $\vec{e}_\varphi=(-\sin\phi, \cos\phi, 0)$. Therefore operator (4) becomes

\begin{equation}
D^2\ \rightarrow\ \partial^2_t-\left(\frac1r\partial_rr\partial_r+\frac{1}{r^2}\partial_\varphi^2+\partial_z^2\right)\ +\ \frac{2i\beta\Theta(r-R)}{r^2}\ \partial_\varphi\ +\ \frac{\beta^2\Theta^2(r-R)}{r^2}
\end{equation} 

here,  and in the rest of this paper, $\beta$ will represent the strength of the background potential

\begin{equation}
\beta\ =\ \frac{e\phi}{2\pi}\ .
\end{equation}
With the ansatz of the separation of the variables the scalar field is transformed into

\begin{equation}
\Phi(x)\ \rightarrow\ \exp(ip_0t-ip_zz+im\varphi)\ \Phi_m(p_0,p_z,r)\ ,
\end{equation}
where $p_\mu$ is the momentum four vector and $m$ is the orbital momentum quantum number.
Combining (4), (8) and (10) the new field equation reads

\begin{equation}
\left(p_0^2-m_e^2-p_z^2-\frac{(m-\beta\Theta(r-R))^2}{r^2}+
\frac 1r\partial_r +\partial_r^2\right)\Phi_m(p_0,p_z,r)\ \ =\ 0 \ .
\end{equation}
or

\begin{equation}
\left(k^2-\frac{(m-\beta\Theta(r-R))^2}{r^2}+
\frac 1r\partial_r +\partial_r^2\right)\Phi_m(k,r)\ =\ 0\ ,
\end{equation}
where $k=\sqrt{p^2_0-m_e^2-p_z^2}$. The solutions of this equation are Bessel and Neumann functions. The kind of function and their coefficients are to be determined by means of physical considerations. We take here the regular solution which in general scattering theory \cite{Taylor} has the following asymptotic behaviour

\begin{equation}
\Phi\ \stackrel{r\rightarrow 0}{\sim} J_m(kr) ,
\end{equation}

\begin{equation}
\Phi\ \stackrel{r\rightarrow \infty}{\sim} \frac 12 (f_m(k)H^{(2)}_{m-\beta}(kr)+f^*_m(k)H^{(1)}_{m-\beta}(kr)) ,
\end{equation}
where $J_m (kr)$ is a Bessel function of the first kind, $H^{(1)}_{m-\beta}(kr)$ and $H^{(2)}_{m-\beta}(kr)$ are Hankel functions of the first and of the second kind and the coefficients $f_m(k)$ and $f^*_m(k)$ are a Jost function and
its complex conjugate respectively. In the case of the delta function profile the regular solution reads

\begin{equation}
\Phi(r)\ =\ J_m (kr)\Theta(R-r)+ \frac 12 (f_m(k)H^{(2)}_{m-\beta}(kr)+f^*_m(k)H^{(1)}_{m-\beta}(kr))\Theta(r-R)\ ,
\end{equation}
then, we can define a field $\Phi^I$ in the region $r<R$ inside the cylinder  

\begin{equation}
\Phi^I_m(k,r)\ =\ J_m (kr)
\end{equation}
which is independent of the strength $\beta$ of the potential, and a field $\Phi^O$ in the region outside the cylinder $rR$ 

\begin{equation}
\Phi^O_m(k,r)\ =\ \frac 12 (f_m(k)H^{(2)}_{m-\beta}(kr)+f^*_m(k)H^{(1)}_{m-\beta}(kr))
\end{equation}
which describes incoming and outgoing cylindrical waves. The conditions for the field at $r=R$ will be discussed later.

\subsection{Ground state energy in terms of the Jost function and normalization condition}

A regularized vacuum energy can be defined as

\begin{equation}
E^{sc}_0\ =\ \frac {\mu^{2s}}{2} \sum \epsilon_{(n,\alpha)}^{1-2s}\ ,
\end{equation}
where the $\epsilon_{(n,\alpha)}$ are the eigenvalues of the Hamiltonian operator associated with (4), $\alpha=\pm 1 $ being the index for the particle-anti-particle degree of freedom, while $n$ includes all oder quantum numbers. $s$ is the regularization parameter to be put to zero after the renormalization and $\mu$ is a mass parameter necessary to maintain the correct dimensions of the energy.
The string is invariant under translations along the $z$ axis, therefore the energy density per unit length of the string is
\begin{equation}
{\cal E}^{sc}\ =\ \frac 12 \mu^{2s}\int^{\infty}_{-\infty}\frac{dp_z}{2\pi}
\sum_{(n,\alpha)} (p_z^2+ \lambda_{(n)}^2)^{1/2-s}\ ,
\end{equation}
where the  $\lambda_{(n)}$ are the eigenvalues of the operator defined in (12)  with $k=\sqrt{p^2_0-m_e^2}$ . We perform the integration over $p_z$ in (19), getting

\begin{equation}
{\cal E}^{sc}\ =\ \frac 14 \mu^{2s}\frac{\Gamma(s-1)}{\sqrt{\pi}\Gamma(s-1/2)}\sum_{(n,\alpha)}(\lambda_{(n)}^2)^{1-s}\ .
\end{equation}
The next step is to transform the summation in (20) into a contour integral. We enclose temporarily the system in a large cylindrical quantization box  imposing some conditions for the field at this boundary, for instance Dirichlet boundary conditions. Then we can express the eigenvalues $\lambda_{(n)}$  by means of the zeros of solution (14), which becomes an exact equation at infinity. The derivative  of the logarithm of the solution will then have poles at $k=\lambda_{(n)}$. We can express (20) through an integral whose contour encloses these poles which lie on the real axis $k$. The deformation of the  contour on the imaginary axis and the dropping of the Minkowski space contribution allows to reach the final form.

\begin{equation}
{\cal E} ^{sc}\ =\ -\frac 12 C_s\ \sum^\infty_{m=-\infty}\int^\infty_{m_e}dk\ (k^2+m_e^2)^{1-s} \partial_k \ln f_m(ik)\ ,
\end{equation}
where $f_m(ik)$ is the Jost function with imaginary argument and $C_s =(1+s(-1+2\ln(2\mu)))/(2\pi)$ is a simple function of the regularization parameter.
The renormalization of (21) is carried out by direct subtraction of its divergent part 

\begin{equation}
{\cal E}_{ren}^{sc}\ =\ {\cal E}^{sc}\ -\ {\cal E}_{div}^{sc}\ .
\end{equation}   
The isolation of the divergent part from the total energy will be performed via heat-kernel expansion as we  will see in a moment. The subtracted part should be added in the classical part of the energy resulting in a renormalization of the classical parameters of the string (in paper \cite{Jou31} this procedure is well explained), however we do not treat the classical energy of the system here but only the vacuum contribution. For the analytical continuation $s\rightarrow 0$  we split ${\cal E}_{ren}^{sc}$ it into a ``finite'' and an ``asymptotic'' part. 

\begin{equation}
{\cal E}_{ren}^{sc}={\cal E}_f^{sc} +{\cal E}_{as}^{sc},
\end{equation}
with

\begin{equation}
{\cal E} _f^{sc}= - \frac 12 C_s \sum_{m=-\infty}^\infty\int^\infty_{m_e} dk [k^2 -m_e^2]^{1-s} \frac{\partial}{\partial k} [\ln f_m(ik)- \ln f^{as}_m(ik)] 
\end{equation}
and
\begin{equation}
{\cal E} _{as}^{sc}= -\frac 12 C_s\sum_{m=-\infty}^\infty \int^\infty_{m_e} dk [k^2 -m_e^2]^{1 -s} 
\frac{\partial}{\partial k}  \ln f^{as}_m(ik)-{\cal E}_{div}^{sc} ,
\end{equation}
where $f^{as}_m$ is a portion of the uniform asymptotic expansion of the Jost function. The number of orders to be included in  $f^{as}_m$ must be sufficient to let the function

\begin{equation}
{\rm  Sub} \ \equiv\ \ln f_m(ik) -  \ln f^{as}_m(ik)
\end{equation}
fall as $m^{-4}$ (or  $k^{-4}$) for $k$ and $m$ equally large, in this case the integral and the summation in (24) converge for $s\rightarrow 0$. To this purpose three orders in the asymptotics are enough. More orders would only give a quicker convergence. The splitting proposed in (23) immediately
permits the analytical continuation $s=0$  in ${\cal E}^{sc}_f$, furthermore  it allows a very quick subtraction of the pole terms in the asymptotic  part (24). ${\cal E}_ {as}^{sc}$ is a finite quantity. For the definition of ${\cal E}_ {div}^{sc}$ we write the sum (18) as 

\begin{equation}
{\cal E} ^{sc}\ =\ \frac 12 \frac{\mu^{2s}}{\Gamma (s-1/2)}\ \int^\infty_0 dt\ t^{s-3/2} K(t)
\end{equation}
where $K(t)$ is the heat kernel related to the Hamilton operator , which can be expanded for small $t$

\begin{equation}
K(t)\ =\ \sum_{(n)} e^{-t \lambda_{(n)}^2}\ \sim\ \frac {e^{-tm_e^2}}{(4\pi t)^{3/2}} \sum_j^\infty A_j t^j\ , \ \ \ j=0,\frac 12 , 1...
\end{equation} 
The $A_j$ are the heat-kernel coefficients related with the Hamiltonian operator. By means of (27) and (28) we can expand the ground state energy in powers of the mass and get

\begin{equation}
{\cal E}^{sc}\ =\ \sum_j
\frac{\mu^{2s}}{32\pi^2}\frac{\Gamma(s+j-2)}{\Gamma(s+1)}m_e^{4-2(s+j)}A_j	
\end{equation}
 in which the divergent contribution can be isolated
\begin{eqnarray} 
{\cal E}_{div}^{sc} &  =  & -\frac{m_e^4}{64 \pi ^2}\left( \frac 1s + \ln
\frac{4\mu ^2 }{m_e^2} - \frac 12\right) A_0\ -\frac{m_e^3}{24\pi^{3/2}}A_{1/2}
\nonumber \\     
                  &     &  +\frac{m_e^2}{32 \pi ^2}\left( \frac 1s +
\ln\frac{4\mu ^2 }{m_e^2} - \ 1\right) A_1 \ + \frac{m_e}{16 \pi^{3/2}}A_{3/2} \nonumber \\ 
                  &     &  -\frac{1}{32 \pi ^2}\left( \frac 1s + \ln \frac{4\mu ^2 }{m_e^2} - 2\right) A_2\ .  
\end{eqnarray} 
In this definition the poles are all contained in the three terms corresponding to the heat kernel coefficients $A_0, A_1, A_2$, however we included in ${\cal E}_{div}^{sc}$ two more terms in order to satisfy a normalization condition, namely that the renormalized ground state energy vanishes for a field of infinite mass

\begin{equation} 
\lim_{m_e \rightarrow \infty }{\cal E}_{ren}^{sc} = 0\ . 
\end{equation} 
This condition fixes a unique value for the vacuum energy of a massive field. It is also necessary to eliminate the arbitrariness of the mass parameter $\mu$. This normalization condition  must of course be changed if one desires to investigate massless fields. In paper \cite{dielectric} the interested reader can find further details and comments about the procedure briefly summarized in this subsection.

\subsection{The Jost function and its asymptotics}
We focus our attention on the Jost function $f_m(k)$ related with the delta function potential of the magnetic string.

 We have defined a field in the outer region $r>R$ and a field in the inner region $r<R$, we now need some matching conditions on the boundary $r=R$. As the delta function is a continuous function, we will require that the field is continuous on the boundary. From this condition and from the field equation (12) it follows directly that the first derivative of the field must be continuous on the boundary:

\begin{equation}
\left\{
\begin{array}{lcl}
\Phi^O(r)|_{r=R} & = &\Phi^I(r)|_{r=R}\\
\partial_r\Phi^O(r)|_{r=R} & = &\partial_r\Phi^I(r)|_{r=R}
\end{array}
\right.
\end{equation}
and with the use of solution (16) and (17)

\begin{equation}
\left\{
\begin{array}{lcl}
J_m(kR) & = & \frac 12 (f_m(k)H^{(2)}_{m-\beta}(kR)+f^*_m(k)H^{(1)}_{m-\beta}(kR)) \\
\partial_r J_m(kr)|_{r=R} & = &\frac 12\partial_r\ (f_m(k)H^{(2)}_{m-\beta}(kr)+f^*_m(k)H^{(1)}_{m-\beta}(kr))|_{r=R}\ \ .
\end{array}
\right.
\end{equation}
Solving for $f_m(k)$ one finds

\begin{equation}
f_m(k)\ =\ \frac{2(\partial_r J_m(kr)|_{r=R}H^{(1)}_{m-\beta}(kR)-\partial_r H^{(1)}_{m-\beta}(kr) |_{r=R} J_m(kR))}{H^{(2)}_{m-\beta}(kR)\partial_r H^{(1)}_{m-\beta}(kr) |_{r=R}-H^{(1)}_{m-\beta}(kR)\partial_r H^{(2)}_{m-\beta}(kr) |_{r=R}}\ ,
\end{equation}
which with the use of the Wronskian determinant for the Hankel functions \cite{Abramowitz} becomes

\begin{equation}
f_m(k)\ =\ -\frac{2\pi kR}{4i}\left[J_m(kR)H^{(1)}_{m-\beta+1}-J_{m+1}(kR)H^{(1)}_{m-\beta}(kR) + \frac{\beta}{kR}J_m(kR)H^{(1)}_{m-\beta}(kR)\right]\ .
\end{equation}
The corresponding Jost function on the imaginary axis can
be written in terms of modified Bessel I and Bessel K functions 

\begin{equation}
f_m(ik)\ =\ i^\beta k R \left[I_m K_{m-\beta+1}+I_{m+1}K_{m-\beta}\right] + i^\beta \beta I_m K_{m-\beta}\ .
\end{equation}
where the arguments $(kR)$ of the Bessel functions are omitted for simplicity. Eq. (36) can be written in a more compact form

\begin{equation}
f_m(ik)\ =\ i^\beta k R\left[I'_m K_{m-\beta}+I_{m}K'_{m-\beta}\right]\ .
\end{equation}
where the prime denotes the derivative  with respect to the argument. This expression holds for positive and for negative values of  $m$. On the contrary the uniform asymptotic expansion, which we need for eq.(24) and (25), is  a different function for positive or for negative $m$. To find it we write the Bessel I and K functions of eq.(36) in the form 

\begin{equation}
I_{m+\alpha}((m+\alpha) z'),\ \ K_{m+\alpha}((m+\alpha) z'), \ \ \ \ \ z'=kR/(m+\alpha),
\end{equation}
where $\alpha$ can be $0$ or $1$ for the Bessel I function and $-\beta$ or $-\beta+1$ for the Bessel K function. Their expansions for large positive orders are well known \cite{Abramowitz}, for instance $K_{m+\alpha}((m+\alpha)z')$ is expanded as

\begin{equation}
K_{m+\alpha}((m+\alpha)z')\ \sim\ \sqrt{\frac{\pi}{2(m+\alpha)}}\frac{e^{-(m+\alpha)\eta '}}{(1+z'^2)^{1/4}}\left\{ 1+\sum_{j=1}^{\infty}(-1)^j \frac{ u_j([1+z'^2]^{-1/2})}{(m+\alpha)^j}\right\}\ ,
\end{equation}
where $\eta=\sqrt{1+z'^2}+\ln(\frac{z'}{1+\sqrt{1+z'^2}})$ and the $u_j(x)$ are the Debye polynomials in the variable $x$. However we are interested in an expansion in powers of the variable $m$ alone: an expansion of the form

\begin{equation}
K_{m+\alpha}((m+\alpha)z')\ \sim\ \sum_n \frac{X_n}{m^n}\ , 
\end{equation}
where the $X_n $ are some coefficients depending on $k,R$ and $\beta$. Therefore we make the substitution $z'\rightarrow z\left(\frac{m}{m+\alpha}\right)$, with $z=KR/m$
in the argument of the Bessel function and we rewrite its  expansion as 

\begin{equation}
\begin{array}{lcl}
K_{m+\alpha}((m+\alpha)z')\ & \sim\ & \sqrt{\frac{\pi}{2(m+\alpha)}}\frac{e^{-(m+\alpha)\eta}}{(1+\left(z\frac{m}{m+\alpha} \right)^2)^{1/4}}\\
                            &       & \left\{ 1+\sum_{j=1}^{\infty}(-1)^j \frac{ u_j([1+\left(z\frac{m}{m+\alpha} \right) ^2]^{-1/2})}{m^j \left(1+\frac{\alpha}{m}\right)^j}\right\}
\end{array}
\end{equation}
with the obvious change for $\eta '$. Then, re-expanding in powers $1/m^n$ we get

\begin{equation}
K_{m+\alpha}(kR)\ \sim\ \sqrt{\frac{\pi}{2m}}\exp \{\sum_{n=-1}^3 m^{-n} S_K(n,\alpha,t)\}\ ,
\end{equation} 
where  $t=(1+z^2)^{-1/2}$ and  the functions $S_K(n,\alpha, t)$ are given explicitly in the Appendix. The corresponding expansion of the Bessel I function in negative powers of $m$ will be 

\begin{equation}
I_{m+\alpha}(kR) \ \sim\ \frac{1}{\sqrt{2\pi m}}\exp \{\sum_{n=-1}^3 m^{-n} S_I(n,\alpha, t)\}\ ,
\end{equation}
where the functions  $S_I(n,\alpha,t)$ are given in the appendix. Inserting these expansions in (36) one finds an asymptotic Jost function valid for positive $m$,  we name it $f^{as+}_m(ik)$. To find the asymptotics for negative  $m$ we must take the Jost function in its form 

\begin{equation}
f_m(ik)\ =\ i^\beta k R\left[I_{-m} K_{-m+\beta-1}+I_{-m-1} K_{-m+\beta}\right] + i^\beta \beta I_{-m} K_{-m+\beta}\ ,
\end{equation}
which is identical to (36) because of  the property of the modified Bessel functions

\begin{equation}
I_l(x)=I_{-l}(x)\ , \ \ \ K_p(x)=I_{-p}(x)\ ,
\end{equation}  
where $l$ is any natural number and $p$ any real number. Then in (44) a large positive index always correspond to a large  negative $m$ and inserting (42) and (43) in (44) we find an asymptotic Jost function valid  for negative $m$, we call it $f^{as-}_m(ik)$. 

For $m=0$, the  asymptotic Jost function can be obtained from (36), using the expansions \cite{Abramowitz}

\begin{equation}
I_{\nu}(z)\ \sim\ \frac{e^z}{\sqrt{2\pi z}}\left\{1-\frac{\mu -1}{8z}+\frac{(\mu -1)(\mu -9)}{2!(8z)^2}-...\right\}\ ,\ \ \ \mu=4\nu^2\ ;
\end{equation}
\begin{equation}
K_{\nu}(z)\ \sim\ \frac{\sqrt{\pi}e^{-z}}{\sqrt{2z}}\left\{1+\frac{\mu -1}{8z}+\frac{(\mu -1)(\mu -9)}{2!(8z)^2}+...\right\}\ ,
\end{equation}
we call this contribution $f_0^{as}(ik)$. The finite and the asymptotic part of the energy  defined in (24),(25) are also split into three contributions: one for positive $m$, one for negative $m$ and one for $m=0$. The positive and negative contributions can be summed up in a single term, but the contribution coming from $m=0$ must be calculated separately and summed just numerically at the end, in fact we have

\begin{eqnarray}
{\cal E} _f^{sc}  & = & - \frac 12 C_s\left(\sum_{m=1}^\infty \int_{m_e}^{\infty} dk\ (k^2+m_e^2) \partial_k (\ln f_m^\pm (ik) -\ln f_m^{as\pm}(ik))\right.\nonumber\\
                  &   & \left. + \int_{m_e}^{\infty} dk\ (k^2+m_e^2) \partial_k (\ln f_0 (ik)- \ln f_0^{as} (ik) )\ ,\right)
\end{eqnarray}
and

\begin{eqnarray}
{\cal E} _{as}^{sc} & = &  - \frac 12 C_s \left( \sum_{m=1}^\infty \int_{m_e}^{\infty} dk\ (k^2+m_e^2)^{1-s} \partial_k \ln f_m^{as\pm}(ik)\right.\nonumber\\
     &   &  \left. +  \int_{m_e}^{\infty} dk\ (k^2+m_e^2)^{1-s} \partial_k \ln f_0^{as}(ik)\ \ -\ {\cal E}_{div}^{sc}\right)\ ,
\end{eqnarray}
where $f_m^\pm= f_m(ik)+f_{-m}(ik)$ and $f^{as\pm}_m(ik)=f^{as+}_m(ik)+f^{as-}_{-m}(ik)$.

 Taking the logarithm of $f^{as+}_m(ik)$ and $f^{as-}_{-m}(ik)$ and re-expanding in powers of $m$ we find the functions needed in the integrands of (48)(49) up to the desired order. We are interested in  $\ln f^{as\pm}_m(ik)$ and $\ln f^{as}_0(ik)$ up to the third order, they read

\begin{equation}
\ln f_0^{as}\ =\ \frac{\beta^2}{2kR}\ ,
\end{equation}
\begin{equation}
\ln f_m^{as\pm}(ik)\ =\ \sum_n^3\sum_t X_{n,j}\frac{t^j}{m^n}\ ,
\end{equation}
where $t=1/(1+(kR/m)^2)^{\frac 12}$ and the nonzero coefficients are

\begin{equation}
\begin{array}{l}
X_{1,1}=\beta^2\ ,\ X_{2,4}=\beta^2/4\ , \  \\
X_{3,3}=\beta^2/24-\beta^4/12\ ,\ X_{3,5}=-\beta^2/2+\beta^4/4\ ,\\ X_{3,7}=\beta^4/16\ .
\end{array}
\end{equation}
As we mentioned above three orders are sufficient for the convergence of ${\cal E} _f^{sc}$. More orders could have been included in definition (50) and (51) to let the sum and the integral in (48) converge more rapidly\footnote{ We would like to stress that with the introduction of asymptotic expansions in our calculation we do {\itshape not} approximate the vacuum energy. The total energy as defined in (22) remains an exact quantity. The uniform asymptotics of the Jost function is just a mathematical tool which permits the analytical continuation $s\rightarrow 0$.}.

\subsection{The asymptotic part of the energy and the heat kernel coefficients}

Having found the Jost function related to the cylindrical delta potential an important part of the calculation is done. We proceed with the analytical simplification of ${\cal E} _{as}^{sc}$. The second term in (49) which we name ${\cal E}_{as0}^{sc}$, can be quickly calculated inserting in it the result (50). We find

\begin{equation}
{\cal E} _{as0}^{sc}\ =\ -\frac{\beta^2m_e}{4 \pi R}\ .
\end{equation}
The first term in (49), which we name here ${\cal E} _{as(m)}^{sc}$ contains the sum over $m$ which can be carried out with the Abel-Plana formula 

\begin{equation}
\sum^\infty_{m=1}F(m)=\int_0^\infty dm F(m)\ -\ \frac 12 F(0) +\ \int_0^\infty\frac{dm}{1-e^{2\pi m}}\frac{F(im)-F(-im)}{i}\ .
\end{equation}
In our case is

\begin{equation}
F(m)=\int_{m_e}^\infty dk (k^2+m_e^2)^{1-s}\partial_k \ln f^{as\pm}_{m}(ik) \ .
\end{equation}
Thus ${\cal E}_{as(m)}^{sc}$ is split into three addends:

\begin{equation}
{\cal E}_{as1}^{sc} \ = -\ \frac 12 C_s \int_0^\infty dm\  \int_{m_e}^\infty dk (k^2+m_e^2)^{1-s}\partial_k \ln f^{as\pm}_{m}(ik)
\end{equation}
\begin{equation}
{\cal E}_{as2}^{sc} \ =\ \frac 14 C_s F(0)\ ,
\end{equation}
\begin{equation}
{\cal E}_{as3}^{sc}\ = -\frac 12 C_s \int_0^\infty\frac{dm}{1-e^{2\pi m}}\frac{F(im)-F(-im)}{i}\ .
\end{equation}
The contributions ${\cal E}_{as1}^{sc} $ and ${\cal E}_{as2}^{sc} $ can be calculated with the formulas given in appendix B. The result is

\begin{equation}
{\cal E}_{as1}^{sc}\ =\ \frac{\beta^2 m_e^2 }{8\pi}\left(\frac 1s +\ln\left(\frac{4\mu^2}{m_e^2}\right) -1\right)\ -\ \frac{\beta^2 m_e}{32 R}\ 
\ ,
\end{equation}
\begin{equation}
{\cal E}_{as2}^{sc}\ =\ \frac{\beta^2 m_e }{4\pi R}\ -\ \frac{\beta^2 }{96\pi m_e R^3}\ +\ \frac{\beta^4 }{48\pi m_e R^3}\ .
\end{equation}
The divergences are all contained in the first term of ${\cal E}_{as1}^{sc}$. The first term of ${\cal E}_{as2}^{sc}$ will cancel with ${\cal E}_{as0}^{sc} $ and  it remains only one term containing a positive power of the mass. This term is the contribution to ${\cal E}^{div}_{sc}$ corresponding to the heat kernel coefficient $A_{3/2}$ (see eq.(30)) and therefore it will be subtracted as well as the pole term.

The last contribution to ${\cal E}_{as}^{sc}$ is ${\cal E}_{as3}^{sc}$, whose calculation demands a little more work. Using the formula displayed in  appendix B to calculate the integral over $k$, we  we find

\begin{equation}
{\cal E}_{as3}^{sc}\ =\ (-1)\frac 12 C_s \sum_{n,j} X_{n,j}\left[-m_e^{2-2s}\Gamma(2-s)\right] \Lambda_{n,j}(m_eR)\ ,
\end{equation}
where  the functions $\Lambda_{n,j}(m_e R)$ are given by

\begin{equation}
\begin{array}{lcl}
\Lambda_{n,j}(x) & = & \frac{\Gamma(s+ j/2 -1)}{\Gamma (j/2)x^j}\left[\int_0^x\frac{dm}{1-e^{2\pi m}}\frac{m^{j-n}\ 2\sin \left[\frac{\pi}{2}(j-n)\right]}{\left[1-\frac{m^2}{x^2}\right]^{s+j/2 -1}} \right.\\
                &   & +\left.\int_x^\infty\frac{dm}{1-e^{s\pi m}}\frac{m^{j-n}\ 2\sin \left[\pi (1-s- n/2)\right]}{\left[\frac{m^2}{x^2}-1\right]^{s+j/2 -1}}   \right]\ .
\end{array}
\end{equation}
We have calculated these functions by partial integration for $n\leq 3$ and $j\leq 7$ and we show them explicitly in the appendix B. By means of those functions and of the coefficients (52)  we find

\begin{eqnarray}
{\cal E}_{as3}^{sc} & = & -\frac{\beta^2}{\pi R^2}\ \int_{m_eR}^{\infty} \frac{dm}{1-e^{2\pi m}} \sqrt{m_e^2 -(m_eR)^2} \nonumber \\
             &   &  +\left( \frac{\beta^2}{24 \pi R^2}-\frac{\beta^4}{12 \pi R^2}\right)\ \int_{m_eR}^{\infty} dm \left(\frac{1}{1-e^{2\pi m}}\frac 1m \right)' \sqrt{m_e^2 -(m_eR)^2} \nonumber \\
             &   & +\left(- \frac{\beta^2}{6\pi R^2}+\frac{\beta^4}{12 \pi R^2}\right)\ \int_{m_eR}^{\infty} dm \left(\left(\frac{m}{1-e^{2\pi m}}\right)'\frac 1m \right)' \sqrt{m_e^2 -(m_eR)^2}\nonumber \\
             &   &  +\frac{\beta^2}{24 \pi R^2}\ \int_{m_eR}^{\infty} dm \left( \left( \left(\frac{m^3}{1-e^{2\pi m}}\right)'\frac 1m \right)' \frac 1m\right)' \sqrt{m_e^2 -(m_eR)^2}\nonumber \\    
             &   &  + \frac{\beta^2 m_e}{16 R}\frac{1}{1-e^{2\pi m_e R}}\ ,
\end{eqnarray}
where the prime in the integrands denotes derivative with respect to $m$. The last term in (63) goes to zero for $m_e\rightarrow \infty$ and therefore must be included in the renormalized energy, however, given its exponential behaviour,  it does not contribute to the heat kernel coefficients. The heat kernel coefficients, which we have calculated up to the coefficient $A_{7/2}$ (including four more orders in $\ln f^{as\pm}(ik)$), read
\begin{equation}
\begin{array}{lclcccl}
A_0 & = & 0                 & , &   A_{1/2} & = & 0 \nonumber \\
A_1 & = &  4\pi\beta^2    & , &   A_{3/2} & = &  \frac{\beta^2 \pi^{3/2}}{2R} \nonumber \\
A_2 & = & 0                & , &   A_{5/2} & = & \frac{[(3\pi-128)\beta^2+(256-18\pi)\beta^4]\pi^{1/2}}{384R^3} \nonumber \\
A_3 & = & 0                & , &  A_{7/2}  & = & \frac{(27\beta^2 -100\beta^4+80\beta^6)\pi^{3/2}}{24576 R^5}\nonumber\ . \\ 
\end{array}
\end{equation}
We perform the subtraction proposed in (22) and we obtain the final result

\begin{eqnarray}
{\cal E}_{as}^{sc} & = & -\frac{\beta^2}{\pi R^2}\  h_1(m_eR)\ +\ \left( \frac{\beta^2}{24 \pi R^2}-\frac{\beta^4}{12 \pi R^2}\right)\ h_2(m_eR)\nonumber \\ 
       &  & +\left( \frac{-\beta^2}{6 \pi R^2}+\frac{\beta^4}{12 \pi R^2}\right)\ h_3(m_eR)\nonumber \\ 
       &  & +\ \frac{\beta^2}{24 \pi R^2}\  h_4(m_eR)\ -\ \frac{\beta^2}{96 \pi m_e R^3}\ +\ \frac{\beta^4}{48 \pi m_e R^3}\nonumber\\
       &  & +\  \frac{\beta^2 m_e}{16 R}\frac{1}{1-e^{2\pi m_e R}}\ .      
\end{eqnarray}
The functions $h_n(x)$ are given by

\begin{eqnarray}
h_1(x) & = & \int_{x}^{\infty} \frac{dm}{1-e^{2\pi m}} \sqrt{m^2 -x^2} \nonumber \\
h_2(x) & = & \int_{x}^{\infty} dm \left(\frac{1}{1-e^{2\pi m}}\frac 1m \right)' \sqrt{m^2 -x^2} \nonumber \\
h_3(x) & = & \int_{x}^{\infty} dm \left(\left(\frac{m}{1-e^{2\pi m}}\right)'\frac 1m \right)' \sqrt{m^2 -x^2} \nonumber \\
h_4(x) & = & \int_{x}^{\infty} dm \left( \left( \left(\frac{m^3}{1-e^{2\pi m}}\right)'\frac 1m \right)' \frac 1m\right)' \sqrt{m^2 -x^2}\ ;
\end{eqnarray}
they are all convergent integrals which can be easily numerically calculated.
 The finite part of the ground state energy given by (48) can be integrated by parts giving

\begin{eqnarray}
{\cal E}_f^{sc} & = & \frac{1}{2\pi}\sum_{m=1}^{\infty}\int_{m_e}^{\infty} dk\ k \left[f^{\pm}_m(ik) - \sum_n^3\sum_t X_{n,j}\frac{t^j}{m^n}\right]\nonumber \\
    &   & +\frac{1}{2\pi}\int_{m_e}^{\infty} dk\ k \left[f_0(ik) - \frac{\beta^2}{2kR}\right]\ ,
\end{eqnarray}
where the coefficient $X_{n,j}$ are given by (52), $t=1/(1+(kR/m)^2)^{\frac 12}$ and
\begin{equation}
f^{\pm}_m(ik)\ =\  k R\left[ i^\beta\left(I'_m K_{m-\beta}+I_{m}K'_{m-\beta}\right)\ +\ i^{-\beta}\left(I'_m K_{m+\beta}+I_{m}K'_{m+\beta}\right)\right]\ ,  
\end{equation}
\begin{equation}
f_0(ik)\ =\ i^\beta k R\left[ I'_0 K_{-\beta}+I_{0}K'_{-\beta}\right]\ .
\end{equation}
Equations(65) and (67) are  to be considered the main analytical result of this paper concerning the scalar field. Their sum gives the total renormalized vacuum energy. The sum will be performed in the numerical part of this paper.

\section{Spinor field in the background of a magnetic string}

\subsection{Solution of the field equation}

The analysis of a spinor field in the background of a cylindrical magnetic field with an arbitrary profile has been performed in \cite{master}. The field equation for a spinor field

\begin{equation}
\Psi(r)\ = \left( \begin{array}{c} g_1(r)\\ g_2(r) \end{array}\right)
\end{equation}
in the background of a translationally invariant potential, with delta function profile is

\begin{equation}
\left( \begin{array}{ll}
p_0-m_e & \partial_r-\frac{m-\beta\Theta(R-r)}{r}\\
-\partial_r-\frac{m+1-\beta\Theta(R-r)}{r} & p_0+m_e
\end{array}\right)
\left( \begin{array}{c} g_1(r)\\ g_2(r) \end{array} \right)\ =\ 0\ .
\end{equation}
The reader is referred to paper \cite{master} for a derivation of this equation. 
Let us find the solution to (71) for one component of the spinor\footnote{We are not interested here in the complete set of solutions to eq.(71), the solution for $g_2(r)$ is sufficient to find  the Jost function of the scattering problem. The decoupled equation for the component $g_1(r)$ is however indispensable and it will be used later.}. The decoupled equation for the component $g_2$ is
\begin{equation}
\left( k^2-\frac{\left[ m-\beta\Theta(R-r)\right]^2}{r^2}+\frac{\beta}{r}\delta(R-r) +\frac 1r\partial_r +\partial_r^2\right)g_2(r)\ =\ 0\ ,
\end{equation}
where $k=\sqrt{p_0^2-m_e^2}$. The solution in the region $r<R$ is

\begin{equation}
g^I_2(r)\ =\ J_m(kr) .
\end{equation}
and in the region $r>R$

\begin{equation}
g^O_2(r)\ =\ \frac 12 \left[f^{spin}_m(k)H^{(2)}_{m-\beta}(kr)+f^{spin^*}_m(k)H^{(1)}_{m-\beta}(kr)\right]\ ,
\end{equation}
here $f^{spin}_m(k)$ and $f^{spin^*}_m(k)$ are the Jost function and its conjugate related to the scattering problem for the spinor field. 
Solutions (73) and (74) have the same form as those found in the scalar case for the inner and outer space. However at $r=R$ the field has not the same form, as we discuss below, owing to the presence of the term $\frac{\beta}{r}\delta(R-r)$ in the field equation (72).

The ground state energy of the spinor field in the background of the magnetic string is
\begin{equation}
E_0\ =\ -\frac{\mu^2}{2}\sum_{n,\alpha,\sigma} \epsilon_{(n,\alpha,\sigma)}^{1-2s}\ ,
\end{equation}
where the minus sign accounts for the change of the statistics, and the $\epsilon_{(n,\alpha,\sigma)}$ are the eigenvalues of the Hamiltonian

\begin{equation}
H\ =\ -i\gamma^0\gamma^l\left(\partial_{x^l}-ieA_l(x)\right)+\gamma^0m_e\ .
\end{equation}
The degree of freedom $\sigma$ accounts for the two independent spin states.
As in the scalar case we calculate the energy for a section of the string. The ground state energy density per unit length of the string in terms of the Jost function is given by

\begin{equation}
{\cal E}^{spin}\ =\ C_s\ \sum^\infty_{m=-\infty}\int^\infty_{m_e}dk\ (k^2-m_e^2)^{1-s} \partial_k \ln f^{spin}_{m}(ik)\ .
\end{equation}
The renormalization scheme is the same we introduced for the scalar case. The expansion of the ground state energy in powers of the mass and the definition of ${\cal E}^{spin}_{div}$ are the same as  in (29), (30), apart from a factor $-1$ coming from the change of the statistics. The heat kernel coefficients will be of course not the same, we call them $B_n$ to distinguish them from those of the scalar problem. The normalization condition (31) remains unchanged. The ground state energy is split into the two parts

\begin{equation}
{\cal E}^{spin}_f= \frac{1}{2\pi} \sum_{m=-\infty}^\infty\int^\infty_{m_e} dk [k^2 -m_e^2] \frac{\partial}{\partial k} [\ln f^{spin}_m(ik)- \ln f^{as-spin}_m(ik)] 
\end{equation}
and
\begin{equation}
{\cal E}^{spin}_{as}= C_s\sum_{m=-\infty}^\infty \int^\infty_{m_e} dk [k^2 -m_e^2]^{1 -s} 
\frac{\partial}{\partial k}  \ln f^{as-spin}_m(ik)-{\cal E}_{div}^{spin}\ ,
\end{equation}
where $f^{as-spin}_m(ik)$ is the uniform asymptotic expansion of the Jost function taken up to the third order in $m$. In (78) the analytical continuation $s\rightarrow 0$ has already been performed, while in ${\cal E}^{spin}_{as}$ it will be performed after the subtraction of the divergent portion. (79) is a finite quantity for $s=0$.
\subsection{Matching conditions and  Jost function}
The matching conditions for the field on the surface of the string are not the same as in the scalar case. More exactly the condition for the first derivative of the field at $r=R$ is different from the one in the scalar case. The field is free inside the magnetic cylinder i.e. independent from $\beta$, then from eq.(71) and (73) we have

\begin{equation} 
k g_1^I(r)+\partial_r g_2^I(r)-\frac mr g_2^I(r)\ =\ 0
\end{equation}
and outside  the cylinder

\begin{equation} 
k g_1^O(r)+\partial_r g_2^O(r)-\frac {m-\beta}{r}g_2^O(r)\ =\ 0\ ,
\end{equation}
the continuity of the field on the boundary is required as in the scalar case:

\begin{equation}
g_2^I(r)|_{r=R}\ =\ g_2^O(r)|_{r=R}\ , \ \ g_1^I(r)|_{r=R}\ =\ g_1^O(r)|_{r=R}\ , 
\end{equation}
therefore, combining (80),(81) and (82) we find

\begin{equation}
(\partial_r g_2^I(r))|_{r=R}-(\partial_r g_2^O(r))|_{r=R}\ =\ \frac{\beta}{r}g_2^I(r)\ .
\end{equation}
Finally, the matching conditions at $r=R$ read
\begin{equation}
\left\{
\begin{array}{l}
g_2^I(r)|_{r=R}\  =\ g_2^O(r)|_{r=R}\\
(\partial_r g_2^I(r))|_{r=R} -(\partial_r g_2^O(r))|_{r=R}\ =\  \frac {\beta}{r} g_2^I(r)|_{r=R}\ .
\end{array}
\right.
\end{equation}
Inserting in (84) solutions (73) and (74) we get the system 
\begin{equation}
\left\{
\begin{array}{lcl}
J_m'- \frac 12 \left[f_m(k)H'^{(2)}_{m-\beta}+f^*_m(k)H'^{(1)}_{m-\beta}\right]\ =\ \frac{\beta}{kR} J_m\\
J_m\ =\ \frac 12 \left[f_m(k)H^{(2)}_{m-\beta}+f^*_m(k)H^{(1)}_{m-\beta}\right]
\end{array}
\right.
\end{equation}
for positive $m$,  and the system
\begin{equation}
\left\{
\begin{array}{lcl}
J_{-m}'- \frac 12 \left[f_m(k)H'^{(2)}_{\beta-m}+f^*_m(k)H'^{(1)}_{\beta-m}\right]\ =\ \frac{\beta}{kR} J_{-m}\\
J_{-m}\ =\ \frac 12 \left[f_m(k)H^{(2)}_{\beta-m}+f^*_m(k)H^{(1)}_{\beta-m}\right]
\end{array}
\right.
\end{equation}
for negative $m$. In (85) and (86) the argument $(kR)$ of the Bessel and of the Hankel functions has been omitted for simplicity and the prime symbol indicates derivative with respect to this  argument. Much in the same way as we did in the scalar case we find the Jost functions on the imaginary axis
\begin{equation}
f_m^{spin+}(ik)\ =\ i^\beta kR \left[I_mK_{m-\beta+1}+I_{m+1}K_{m-\beta}\right], \ \ m>0  
\end{equation}
\begin{equation}
f_m^{spin-}(ik)\ =\ i^{-\beta} kR \left[I_mK_{m-\beta+1}+I_{m+1}K_{m-\beta}\right], \ \ m<0\ .  
\end{equation}
The two Jost functions are identical apart form the sign on the exponent of the imaginary factor. However  formula (77) for the ground state energy contains the logarithm of the Jost function and the derivative with respect to $k$, then the imaginary factor $i^{\pm\beta}$ which is independent of $k$ will not contribute to the energy. 

For the calculation of the asymptotic Jost function we found more convenient to use instead of $m$ the expansion parameter $\nu$ given by
\begin{equation}
\nu\ =\ \left\{ \begin{array}{rcl}
m+ 1/2  & {rm for} & m=0,1,2...\\
-m- 1/2 & {rm for} &  m=-1,-2,... 
\end{array}
\right.
\end{equation} 
with $\nu=\frac 12, \frac 32,...$ in both cases. Then the Jost functions (87) and (88) become
\begin{equation}
f_{\nu}^+(ik)\ =\ i^\beta kR \left[I_{\nu+\frac 12} K_{\nu-\frac 12-\beta}+I_{\nu-\frac 12}K_{\nu+\frac 12 -\beta}\right],\ \ m\geq 0  
\end{equation}
which can be expanded for large positive $m$
\begin{equation}
f_{\nu}^-(ik)\ =\ i^{-\beta} kR \left[I_{\nu+\frac 12} K_{\nu-\frac 12 +\beta}+I_{\nu-\frac 12}K_{\nu+\frac 12 + \beta}\right]\ ,\ \ m<0  
\end{equation}
which can be expanded for large negative $m$.
The asymptotic expansions of the Bessel I and K functions for $\nu$ and $k$ equally large are given by
\begin{equation}
I_{\nu+\alpha}\ \sim\ \frac{1}{\sqrt{2\pi\nu}}\exp \left\{ \sum_{n=-1}x^n S_I(n,\alpha,t) \right\}\ ,
\end{equation}
\begin{equation}
K_{\nu+\alpha\pm \beta}\ \sim\ \frac{\sqrt{\pi}}{\sqrt{2\nu}}\exp \left\{ \sum_{n=-1}x^n S_K(n,\alpha,t) \right\}\ ,
\end{equation}
where $x\equiv 1/\nu$, the functions  $S_I(n,\alpha,t)$, $S_K(n,\alpha,t)$  are the same as in the scalar case  and $\alpha$ takes the values $\pm \frac 12$ for the Bessel I function and $\pm\frac 12\pm\beta$ for the Bessel K function. From these formulas the logarithm of the asymptotic Jost function can be easily calculated up to the third order and we define
\begin{equation}
\ln f^{as-spin}_{\nu}(ik)\ =\ \sum_{j,n}^3 Y_{j,n}\frac{t^j}{\nu^n}\ ,
\end{equation}
where  $t=1/(1+(kR/\nu)^2)^{\frac 12}$ and the nonzero coefficients are
\begin{equation}
\begin{array}{l}
Y_{1,1}=\beta^2\ ,\ Y_{2,2}=-\beta^2/4\ ,\ Y_{2,4}=\beta^2/4\ , \  \\
Y_{3,3}=\beta^2/6-\beta^4/12\ ,\ Y_{3,5}=-7\beta^2/8+\beta^4/4\ ,\\ Y_{3,7}=5\beta^2/8\ .
\end{array}
\end{equation}

\subsection{The asymptotic and the finite part of the energy}

The asymptotic part of the energy can be written, using result (94), as
\begin{equation}
{\cal E}^{spin}_{as}=  C_s\sum_{\nu=\frac 12}^\infty \int^\infty_{m_e} dk [k^2 -m_e^2]^{1 -s} 
\frac{\partial}{\partial k}\sum_{j,n}^3 Y_{j,n}\frac{t^j}{\nu^n}  -{\cal E}_{div}^{spin} ,
\end{equation}
we calculate the sum over $\nu$ with the help of the Abel Plana formula for half integer variables which can be found in the appendix. The case $m=0$ i.e. $\nu=1/2$ does not need to be treated separately. We have only the two contributions
\begin{equation}
{\cal E}^{spin}_{as1}\ =\ \frac{\beta^2m_e^2}{4\pi}\left(\frac 1s +\ln\left(\frac{4\mu^2}{m_e^2}\right) -1\right)\-\ -\frac{\beta^2m_e}{16R} 
\end{equation}
and
\begin{equation} 
{\cal E}^{spin}_{as2}\ =\ C_s \sum_{n,j}^{3,7}Y_{n,j}(-m_e^{2-2s}\Gamma(2-s))\Sigma_{n,j}(m_eR)\ .
\end{equation}
The functions $\Sigma_{n,j}(x)$ are given in the appendix. They correspond to the functions $\Sigma_{n,j}(x)$  found in paper \cite{master} for a generic smooth background potential.
The only pole term is contained in ${\cal E}^{spin}_{as1}$ and the term proportional to $m_e$ will be subtracted as well as the pole term, thus (97) cancels  completely with the subtraction of ${\cal E}_{div}^{spin}$. The heat kernel coefficients which we have calculated up to the coefficient $B_4$, read

\begin{equation}
\begin{array}{lclcccl}
B_0 & = & 0                 & , &   B_{1/2} & = & 0 \nonumber \\
B_1 & = &  8\pi\beta^2    & , &   B_{3/2} & = &  -\beta^2 \pi^{3/2}/R \nonumber \\
B_2 & = & 0                & , &   B_{5/2} & = & \frac{(3\beta^2+2\beta^4)\pi^{3/2}}{64R^3} \nonumber \\
B_3 & = & 0                & , &  B_{7/2}  & = & -\frac{(135\beta^2 -68\beta^4+16\beta^6)\pi^{3/2}}{12288 R^5}\nonumber \\
B_4 & = &  \frac{5\beta^8}{1281 R^6}  & . &  &  & 
\end{array}
\end{equation}
After the subtraction of ${\cal E}_{div}^{spin}$ the asymptotic part of the energy reads
\begin{eqnarray}
{\cal E}^{spin}_{as} & = & \frac{2\beta^2}{\pi R^2}\  q_1(m_R)\ +\ \left( -\frac{\beta^2}{3 \pi R^2}+\frac{\beta^4}{6 \pi R^2}\right)\ q_2(m_eR)\nonumber \\ 
       &  & +\left( \frac{7\beta^2}{12 \pi R^2}+\frac{\beta^4}{6 \pi R^2}\right)\ q_3(m_eR)\ -\ \frac{\beta^2}{12 \pi R^2}\  q_4(m_eR)\ ,      
\end{eqnarray}
the functions $q_n(x)$ are
\begin{eqnarray}
q_1(x) & = & \int_{x}^{\infty} \frac{d\nu}{1+e^{2\pi \nu}} \sqrt{\nu^2 -x^2} \nonumber \\
q_2(x) & = & \int_{x}^{\infty} d\nu \left(\frac{1}{1+e^{2\pi \nu}}\frac {1}{\nu} \right)' \sqrt{\nu^2 -x^2} \nonumber \\
q_3(x) & = & \int_{x}^{\infty} d\nu \left(\left(\frac{\nu}{1+e^{2\pi \nu}}\right)'\frac {1}{\nu} \right)' \sqrt{\nu^2 -x^2} \nonumber \\
q_4(x) & = & \int_{x}^{\infty} d\nu \left( \left( \left(\frac{\nu^3}{1+e^{2\pi \nu}}\right)'\frac {1}{\nu} \right)' \frac {1}{\nu}\right)' \sqrt{\nu^2 -x^2}\ .
\end{eqnarray}
It is interesting to note how in both scalar and spinor cases we found the final expression for the asymptotic part of the energy to depend only on even powers of $\beta$. This was to expect for physical reasons, infact, inverting the direction of the magnetic flux $\phi$  the ground state energy should not change.
 
The finite part of the energy  ${\cal E}^{spin}_f$ can be hardly analytically simplified. We can only integrate expression (78) by parts to obtain a final form which is suitable for the numerical calculation:
\begin{eqnarray}
{\cal E}^{spin}_f & = & -\frac{1}{\pi}\sum_{\nu=\frac 12}^{\infty}\int_{m_e}^{\infty} dk\ k \left[\ln f^{\pm}_{\nu}(ik) - \sum_{n,j}^{3,7} Y_{n,j}\frac{t^j}{\nu^n}\right]\ ,
\end{eqnarray}
where
\begin{eqnarray}
f^{\pm}_{\nu}(ik) & = & kR \left\{ \left(I_{\nu+\frac 12} K_{\nu-\frac 12-\beta}+I_{\nu-\frac 12}K_{\nu+\frac 12 -\beta}\right)\right.\nonumber \\
                  &    &\ \ \   \left. +\left(I_{\nu+\frac 12} K_{\nu-\frac 12 +\beta}+I_{\nu-\frac 12}K_{\nu+\frac 12 + \beta}\right) \right\}
\end{eqnarray}\ .

\section{Numerical evaluations}

In this section we show some graphics of ${\cal E}_{as}$,  ${\cal E}_{f}$ and of the complete renormalized vacuum energy ${\cal E}_{ren}$ as a function of the radius of the string, for the scalar and for the spinor field. We calculate also the asymptotic behavior of  ${\cal E}_{as}$ and ${\cal E}_f$ for large and for small $R$. Since we want to study here only the dependence  on $R$ and on $\beta$, we set $m_e=1$ for all the calculations of this section.  

\subsection{Scalar field}
As a first step we rewrite  ${\cal E}^{sc}_{as}$ in a form in which the dependence on the relevant parameters is more explicit:
\begin{equation}
{\cal E}^{sc}_{as}\ =\ \frac{1}{\pi R^2}\left[\beta^2 g_1(m_eR)\ +\ \beta^4 g_2(m_e R)\right]\ ,
\end{equation} 
with
\begin{eqnarray}
g_1(x) & = & \left(-h_1(x)+\frac{1}{24}h_2(x)-\frac 16 h_3(x) +\frac{1}{24}h_4(x)-\frac{1}{96x} + \frac{x}{16(1-e^{2\pi x})}\right)\ ,\nonumber\\
g_2(x) & = & \left(-\frac{1}{12}h_2(x)+\frac{1}{12}h_3(x)+\frac{1}{48x}\right)\ .
\end{eqnarray}
the asymptotic behaviour of the $h_n(x)$ functions for $x\rightarrow 0$ is found to be
\begin{eqnarray}
h_1(x) & \sim & -1/24\ +{\cal O}(x)\ ,\nonumber\\
h_2(x) & \sim & \frac{1}{4x}+\frac 12 \ln x -0.0575\ + {\cal O}(x)\ ,\nonumber\\
h_3(x) & \sim & \frac 12 \ln x +0.442\ +{\cal O} (x)\ ,\nonumber\\
h_4(x) & \sim & \frac 32\ln x + 1.826\ +{\cal O} (x)\ ;
\end{eqnarray}
the corresponding behaviour of the functions $g_n(x)$ is
\begin{eqnarray}
g_1(x) & \sim & 0.0317 +\ {\cal O}(x)\ ,\nonumber\\
g_2(x) & \sim & 0.0417   +\ {\cal O}(x) \ .
\end{eqnarray}
The logarithmic contributions have cancelled. This was to expect also from the vanishing of the heat kernel coefficient $A_2$. ${\cal E}^{sc}_{as}$ is proportional to  $R^{-2}$ for $R\rightarrow 0$ and for an arbitrary $\beta$. For $R\rightarrow \infty$ all the $h_n(x)$ functions fall exponentially and so does ${\cal E}^{sc}_{as}$.
 
The finite part ${\cal E}^{sc}_f$ is also proportional to $R^{-2}$ in the limit $R\rightarrow 0$, it can be seen substituting $k\rightarrow k/R$ in the integrands of expression (67)
\begin{eqnarray}
{\cal E}^{sc}_f & = & \frac{1}{2\pi R^2}\sum_{m=1}^{\infty}\int_{m_eR}^{\infty} dk k \left[f^{\pm}_m(ik)|_{k\rightarrow k/R} - \sum_n^3\sum_t X_{n,j}\frac{t^j}{m^n}\right]\nonumber \\
    &   & +\frac{1}{2\pi R^2}\int_{m_e}^{\infty} dk\ k \left[f_0(ik)|_{k\rightarrow k/R} - \frac{\beta^2}{2k}\right]\ ,
\end{eqnarray}
where the Jost functions are now independent of $R$. We name the first addend in (108) ${\cal E}^{sc}_{fm}$ and second addend ${\cal E}^{sc}_{f0}$, in the plots we will display them separately. For large $R$ we found numerically ${\cal E}^{sc}_f\sim R^{-3}$, which is in agreement with the heat kernel coefficient $A_{5/2}$ shown in (64), in fact the first non vanishing heat kernel coefficient after $A_2$ determines the behaviour of the renormalized energy for $R\rightarrow \infty$. Below we show the plots of 
all the contributions to the renormalized ground state energy. The functions $g1(x)$ and  $g1(x)$ are shown as well. Each contribution has been multiplied by $R^2$ so that all the curves take a finite value at $R=0$. We found necessary to sum up to $20$ in the parameter $m$ an to integrate up to $1000$ in the $k$ variable in order to obtain reliable plots. All the calculations where performed with computer programming relying on a precision of $34$ digits.

\subsection{Spinor field}
The asymptotic part of the energy is rewritten in the form
\begin{equation}
{\cal E}^{spin}_{as}\ =\ \frac{1}{12\pi R^2}\left[\beta^2 e_1(m_eR)\ +\ \beta^4 e_2(m_e R)\right]\ ,
\end{equation}
with
\begin{eqnarray}
e_1(x) & = & \left(24q_1(x)-4q_2(x)+7 q_3(x)-q_4(x)\right)\ ,\nonumber\\
e_2(x) & = & 2\left(q_2(x)-q_3(x)\right)\ .
\end{eqnarray}
The asymptotic behaviour of the $q_n(x)$ functions for $x\rightarrow\ 0$ is
\begin{eqnarray}
q_1(x) & \sim & 1/48\ + {\cal O}(x)\ ,\nonumber\\
q_2(x) & \sim & \frac 12 \ln x +0.635\ +{\cal O}(x)\ ,\nonumber\\
q_3(x) & \sim & \frac 12 \ln x +1.135\ +{\cal O}(x)\ ,\nonumber\\
q_4(x) & \sim & \frac 32\ln x + 3.906\ +{\cal O}(x)
\end{eqnarray}
and the corresponding behaviour of the $e_n(x)$ functions is
\begin{eqnarray}
e_1(x) & \sim & 2 +\ {\cal O}(x)\ ,\nonumber\\
e_2(x) & \sim & -1 +\ {\cal O}(x) \ ;
\end{eqnarray}
therefore ${\cal E}^{spin}_{as}$ behaves like $R^{-2}$ for small values of $R$.
For $R\rightarrow\infty$ ${\cal E}^{spin}_{as}$ falls like $e^{-R}$.

The finite part of the energy can be rewritten as
\begin{eqnarray}
{\cal E}^{spin}_f & = & -\frac{1}{\pi R^2}\sum_{\nu=\frac 12}^{\infty}\int_{m_eR}^{\infty} dk\ k \left[\ln f^{\pm}_{\nu}(ik)|_{k\rightarrow k/R} - \sum_{n,j}^{3,7} Y_{n,j}\frac{t^j}{\nu^n}\right]\ ,
\end{eqnarray}
which is proportional to $R^{-2}$ for $R\rightarrow 0$ and to $R^{-3}$ for $R\rightarrow \infty$ in agreement with the heat kernel coefficient $B_{5/2}$. We give the plots of the functions $e_1(x)$, $e_2(x)$ of ${\cal E}^{spin}_{as}$, ${\cal E}^{spin}_f$ and of ${\cal E}^{spin}_{ren}$, for small and for large values of the potential strength. The same remarks which we made in the scalar case about the numerical limits of summation and integration and about the precision employed in the computer graphics are valid here.  

\section{Conclusions}
In this paper we have carried out a complete calculation of the vacuum energy of two different fields in the background of a magnetic string with delta function profile. The renormalized vacuum energy is given in terms of convergent integrals (eq.(65) and (67) for the scalar field, eq.(100) and (102) for the spinor field). A first remark can be made about the vanishing of the heat kernel coefficient $A_2$ in both scalar and spinor cases. This coefficient, contributing to ${\cal E}_{div}$, is not zero for a generic background potential. The vanishing of $A_2$ is also observed in a dielectric spherical shell with a squared profile \cite{dielectric}. It could be argued that singular profiles possess less ultraviolet divergences than smooth profiles. This statement is also confirmed by the heat kernel coefficients calculated in \cite{heat kernels}. 

The dependence of the sign of the energy on the radius $R$ of the string and on the potential strength $\beta$ is non trivial. In the scalar case the energy is negative only for large values of the potential strength, while for $\beta$ smaller than one, the energy shows a positive maximum (Fig. 5). In the spinor case we have almost an  opposite situation (Fig. 10): in the region $R<1$ the energy shows a negative minimum for $\beta < 1$, while it is  positive for $\beta > 1 $. However, when $R$ becomes large the vacuum energy shows the same behaviour for the scalar and for the spinor field: it is negative for large $\beta$ and positive for small $\beta$. This strong dependence on the parameter $\beta $ was not observed in paper \cite{master}, where an homogeneous field inside the flux tube was investigated. In fact the most relevant result of our calculation is that the energy numerically shows a dependence on $\beta^{4}$ for large $\beta$. In \cite{master} the contributions proportional to $\beta^{4}$ cancelled and the parts proportional to  $\beta^{2}$  dominated the renormalized energy. The proportionality  ${\cal E}_{ren}\sim \beta^{4}$ opens the possibility that the inhomogeneity of the magnetic field could render the vacuum energy  larger than the classical energy of the string, when the boundary becomes sufficiently hard. The total energy of the system could, then, be dominated by the quantum contribution, in fact we have
\begin{equation}
E_{TOT}\ \sim \ \underbrace{\frac{\beta^2 {\cal A}}{\alpha R^2}}_{E_{class}}\ +\ \underbrace{\frac{\beta^2 {\cal B} +\beta^4 {\cal C}}{R^2}}_{E_{vacuum}}\ ,
\end{equation}
where ${\cal A}$, ${\cal B}$ and ${\cal C}$ are numbers and $\alpha $ is the fine structure constant.
However in the model studied here, the profile of the potential contains a delta function and the classical energy is infinite. It would be interesting to study an inhomogeneous magnetic field  which does not contain singularities in order to have a finite classical energy and possibly a vacuum energy depending on $\beta^{4}$. Unfortunately with more realistic potentials the calculations become more difficult. A feasible model would be that of a cylindrical shell with finite thickness and with a profile given by a finite height box. In this case the Jost function would be expressed in terms of Bessel functions and hypergeometric functions. This problem is left for future investigation.

\section{Acknowledgments} 

I thank M. Bordag for advice.

\section{Appendix A: List of the functions used for the asymptotic expansions of the modified Bessel functions}

The functions $S_I(n,\alpha,t)$ and  $S_k(n,\alpha,t)$  used in (42)(43) and (92)are
\begin{eqnarray}
S_I(-1,\alpha,t) & = & t^{-1} + \frac 12 \ln\left(\frac{1 - t}{1 + t}\right)\ ,\nonumber \\
 S_I(0,\alpha,t) & = & \frac 12\ln t -\frac{\alpha}{2} \ln\left(\frac{1 + t}{1 - t}\right)\ ,\nonumber \\
S_I(1, \alpha,t)& = & -\frac {t}{24} (-3 + 12\alpha^2 + 12\alpha t + 5 t^2)\ ,\nonumber \\
S_I(2, \alpha,t) & = & \frac{t^2}{48} [8\alpha^3 t + 12\alpha^2 (-1 + 2 t^2) + 
       \alpha(-26 t + 30 t^3) + 3 (1 - 6 t^2 + 5 t^4)]\ ,\nonumber \\
S_I(3, \alpha,t) & = & \frac {1}{128}(((25 - 104\alpha^2 + 16\alpha^4) t^3)/3 + 
      16 \alpha (-7 + 4\alpha^2) t^4\nonumber\\
               &   & - (531/5 - 224\alpha^2 + 16\alpha^4)t^5 - (32\alpha (-33 + 8\alpha^2) t^6)/3\nonumber \\
               &   & - (-221 + 200\alpha^2) t^7 - 240\alpha t^8 - (1105 t^9)/9)\ ;
\end{eqnarray}

\begin{eqnarray}
S_K(-1,\alpha,t) & = & -t^{-1} -\frac 12 \ln\left(\frac{1 - t}{1 + t}\right)\ ,\nonumber \\
 S_K(0,\alpha,t) & = & \frac 12\ln t +\frac{\alpha}{2} \ln\left(\frac{1 + t}{1 - t}\right)\ ,\nonumber \\
S_K(1, \alpha,t)& = & -\frac {t}{24} (-3 + 12\alpha^2 + 12\alpha t + 5 t^2)\ ,\nonumber \\
S_K(2, \alpha,t) & = & \frac{t^2}{48} [-8\alpha^3 t + 12\alpha^2 (-1 + 2 t^2) + 
       \alpha(-26 t + 30 t^3) + 3 (1 - 6 t^2 + 5 t^4)]\ ,\nonumber \\
S_K(3, \alpha,t) & = & -\frac {1}{128}((25 - 104\alpha^2 + 16\alpha^4) t^3)/3 + 
      16 \alpha (-7 + 4\alpha^2) t^4\nonumber\\
               &   & - (-531/5 + 224\alpha^2 - 16\alpha^4)t^5- (32\alpha (-33 + 8\alpha^2) t^6)/3\nonumber\\
               &   & - (221 - 200\alpha^2) t^7 - 240\alpha t^8 + (1105 t^9)/9\ ,
\end{eqnarray}
where $t=1/(1+(kR/m)^2)^{\frac 12}$ in the scalar case and $t=1/(1+(kR/\nu)^2)^{\frac 12}$ in the spinor case.

\section{Appendix B: Calculation of the integrals}

The transformation of the sum in (96) into an integral has been done with the Abel-Plana formula for the summation over half integer numbers \cite{Mostepanenko}
\begin{equation}
\sum_{m=0}^\infty F(m+\frac 12)\ =\ \int_0^\infty d\nu\ F(\nu)\ +\ \int_0^\infty\frac{d\nu}{1+e^{2\pi\nu}}\frac{F(i\nu)-F(-i\nu)}{i}\ .
\end{equation}
The following formulas, taken from \cite{master}, have been used for the integration over $m$ and  $k$ in ${\cal E}^{sc}_{as1}$,  ${\cal E}_{as2}^{sc}$ and in ${\cal E}_{as(1)}^{spin}$ (eq.(56),(57) and (97))
\begin{equation}
\int_0^\infty dm\ \int_{m_e}^\infty dk\ (k^2-m_e^2)^{1-s}\partial_k\frac{t^j}{m^n}=-\frac{m_e^{2-2s}}{2}\frac{\Gamma(2-s)\Gamma\left(\frac{1+j-n}{2}\right)\Gamma(s+\frac{n-3}{2})}{(Rm_e)^{n-1}\Gamma(j/2)}\ ,
\end{equation}
where $t=1/(1+(kR/m)^2)^{\frac 12}$ in the scalar case and $t=1/(1+(kR/\nu)^2)^{\frac 12}$ in the spinor case. For the calculation of the integrals over $k$ in $E_{as}^{(1)}$ (eq. (58)) and  in $E_{as(2)}^{spin}$ (eq.(98)) we used
\begin{equation}
\int_{m_e}^\infty dk\ (k^2-m_e^2)^{1-s}\partial_k\frac{t^j}{m^n}=-\frac{m_e^{2-2s}}{2}\frac{\Gamma(2-s)\Gamma\left(\frac{1+j-n}{2}\right)\Gamma(s+\frac{n-3}{2})}{(Rm_e)^{n-1}\Gamma(j/2)}\ ,
\end{equation}
from which one can derive expression (61) and the analogous expression for the spinor field. The functions $\Lambda_{n,j}(x)$ and $\Sigma_{n,j}(x)$ up to $n=3,j=7$ read 
\begin{eqnarray}
\Lambda_{1,1}(x) & = & -\frac{4}{x^2}\int_x^\infty \frac{dm}{1-e^{2\pi m}} \sqrt{m^2 -x^2} \nonumber \\
\Lambda_{2,4}(x) & = & -\frac{\pi}{x}\frac{1}{1-e^{2\pi x}}\nonumber\\
\Lambda_{3,3}(x) & = & -\frac{4}{x^2}\int_{x}^{\infty} dm \left(\frac{1}{1-e^{2\pi m}}\frac 1m \right)' \sqrt{m^2 -x^2} \nonumber \\
\Lambda_{3,5}(x) & = & -\frac{4}{x^2}\int_{x}^{\infty} dm\ \frac 13 \left(\left(\frac{m}{1-e^{2\pi m}}\right)'\frac 1m \right)' \sqrt{m^2 -x^2} \nonumber \\
\Lambda_{3,7}(x) & = & -\frac{4}{x^2}\int_{x}^{\infty} dm \ \frac{1}{15}\left( \left( \left(\frac{m^3}{1-e^{2\pi m}}\right)'\frac 1m \right)' \frac 1m\right)' \sqrt{m^2 -x^2}\ ;
\end{eqnarray}
\begin{eqnarray}
\Sigma_{1,1}(x) & = & -\frac{4}{x^2}\int_x^\infty \frac{d\nu}{1+e^{2\pi \nu}} \sqrt{\nu^2 -x^2} \nonumber \\
\Sigma_{3,3}(x) & = & -\frac{4}{x^2}\int_{x}^{\infty} d\nu \left(\frac{1}{1+e^{2\pi \nu}}\frac 1\nu \right)' \sqrt{\nu^2 -x^2} \nonumber \\
\Sigma_{3,5}(x) & = & -\frac{4}{x^2}\int_{x}^{\infty} d\nu\ \frac 13 \left(\left(\frac{\nu}{1+e^{2\pi m}}\right)'\frac 1\nu \right)' \sqrt{\nu^2 -x^2} \nonumber \\
\Sigma_{3,7}(x) & = &  -\frac{4}{x^2}\int_{x}^{\infty} d\nu \ \frac{1}{15}\left( \left( \left(\frac{\nu^3}{1+e^{2\pi \nu}}\right)'\frac 1\nu \right)' \frac 1\nu \right)' \sqrt{\nu^2 -x^2}\ ;
\end{eqnarray}

\vfill \eject

\begin{figure}[ht]\unitlength1cm
\begin{picture}(6,6)
\put(-0.5,0){\epsfig{file=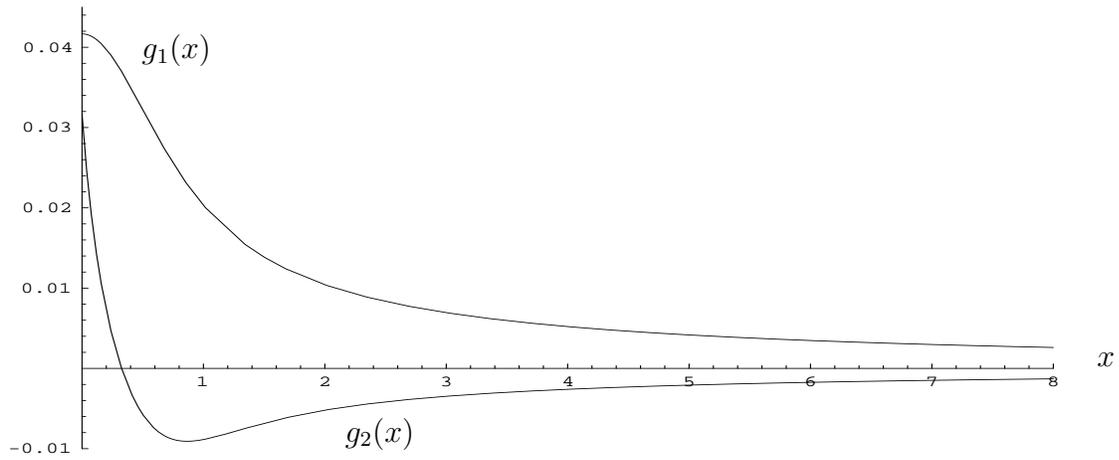,width=14cm,height=6cm}}
\put(1.3,5.3){$g_1(x)$}
\put(14,1.2){$x$}
\put(4,0.2){$g_2(x)$}
\end{picture}
\caption{ Scalar field. The functions $g_1(x)$ and $g_2(x)$ contributing to the asymptotic part of the ground state energy.} 
\end{figure}

\begin{figure}[ht]\unitlength1cm
\begin{picture}(6,6)
\put(-0.5,0){\epsfig{file=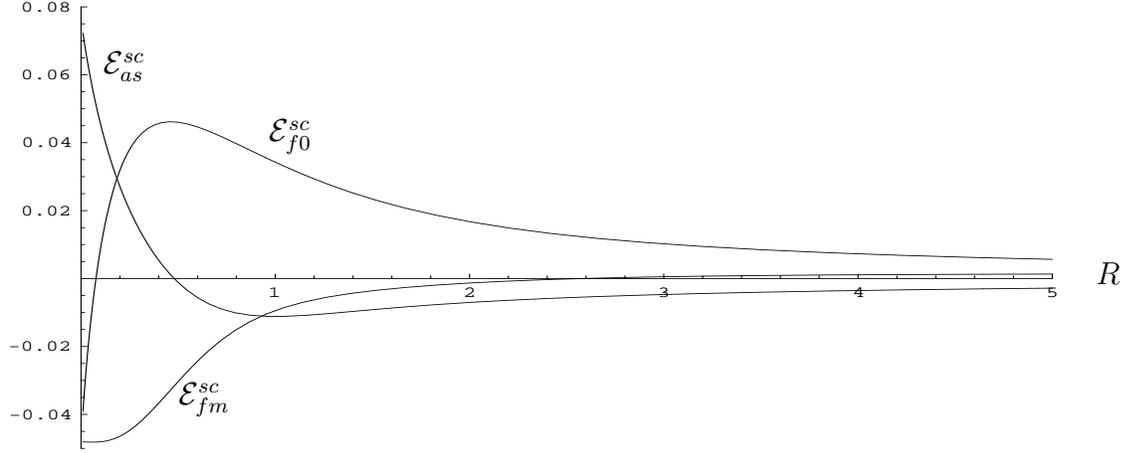,width=14cm,height=6cm}}
\put(3.0,4.1){${\cal E}^{sc}_{f0}$}
\put(1.8,0.6){${\cal E}^{sc}_{fm}$}
\put(0.8,5.0){${\cal E}^{sc}_{as}$}
\put(14,2.2){$R$}
\end{picture}
\caption{ Scalar field. The  three  contributions to the renormalized vacuum energy multiplied by $R^{2}\cdot \beta^{-4}$, for $\beta =0.4$. The curve ${\cal E}^{sc}_{f0}$ shows the contribution to the finite part of the energy coming from the orbital momentum $m=0$, while ${\cal E}^{sc}_{fm}$ displays the contribution coming from the sum of all other $m$'s.} 
\end{figure}

\begin{figure}[ht]\unitlength1cm
\begin{picture}(6,6)
\put(-0.5,0){\epsfig{file=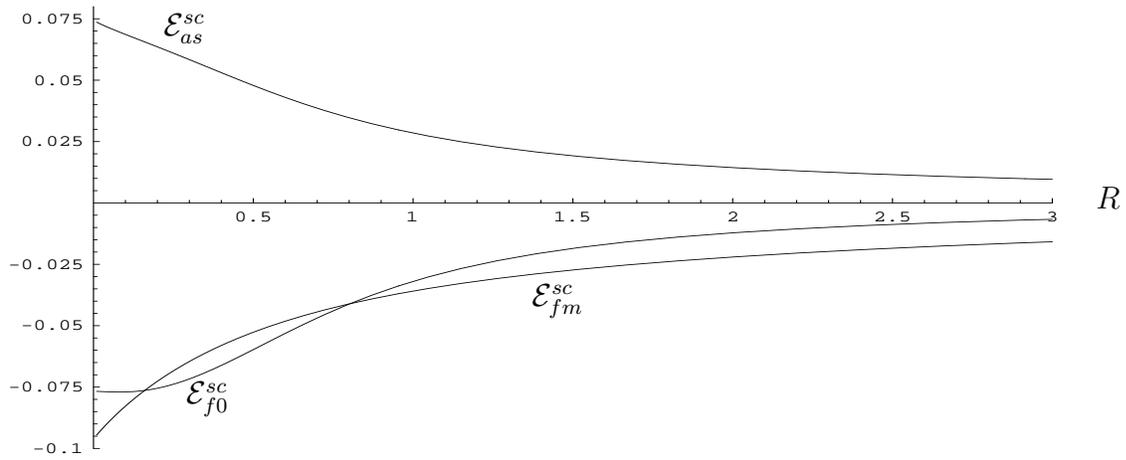,width=14cm,height=6cm}}
\put(1.6,5.6){${\cal E}^{sc}_{as}$}
\put(6.5,2.0){${\cal E}^{sc}_{fm}$}
\put(1.9,0.7){${\cal E}^{sc}_{f0}$}
\put(14,3.3){$R$}
\end{picture}
\caption{ Scalar field. ${\cal E}^{sc}_{as}$, ${\cal E}^{sc}_{fm}$ and ${\cal E}^{sc}_{f0}$ multiplied by $R^{2}\cdot \beta^{-4}$, for $\beta =2.2$.} 
\end{figure}

\begin{figure}[ht]\unitlength1cm
\begin{picture}(6,6)
\put(-0.5,0){\epsfig{file=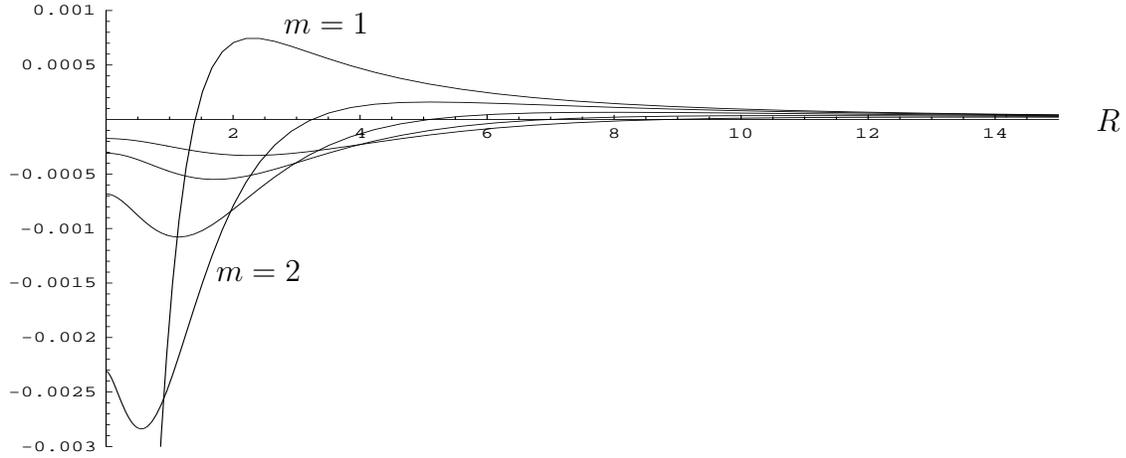,width=14cm,height=6cm}}
\put(2.3,2.3){$m=2$}
\put(3.2,5.6){$m=1$}
\put(14,4.3){$R$}
\end{picture}
\caption{ Scalar field. The contributions $m=1,2...5$ to the finite part of the energy multiplied by $R^{2}\cdot \beta^{-4}$, for $\beta =2.2$.} 
\end{figure}

\begin{figure}[ht]\unitlength1cm
\begin{picture}(6,6)
\put(-0.5,0){\epsfig{file=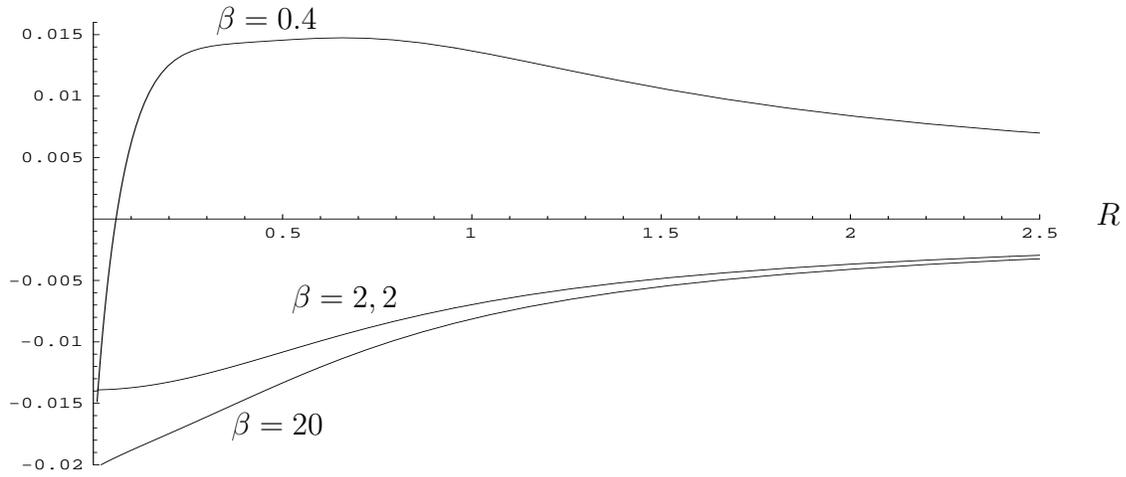,width=14cm,height=6cm}}
\put(2.5,0.5){$\beta=20$}
\put(2.3,5.9){$\beta=0.4$}
\put(3.3,2.2){$\beta=2,2$}
\put(14,3.3){$R$}
\end{picture}
\caption{ Scalar field. The complete renormalized vacuum energy  ${\cal E}_{ren}^{sc}(R)$ multiplied by $R^{2}\cdot \beta^{-4}$, for different values of strength of the potential.} 
\end{figure}

\begin{figure}[ht]\unitlength1cm
\begin{picture}(6,6)
\put(-0.5,0){\epsfig{file=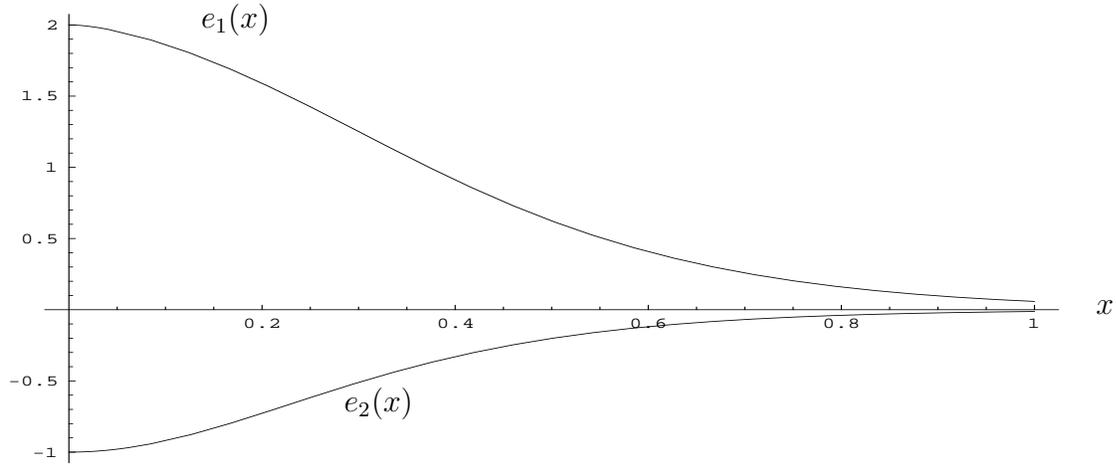,width=14cm,height=6cm}}
\put(2.1,5.8){$e_1(x)$}
\put(14,2){$x$}
\put(4,0.7){$e_2(x)$}
\end{picture}
\caption{ Spinor field. The functions $e_1(x)$ and $e_2(x)$ contributing to the asymptotic part of the ground state energy.} 
\end{figure}

\begin{figure}[ht]\unitlength1cm
\begin{picture}(6,6)
\put(-0.5,0){\epsfig{file=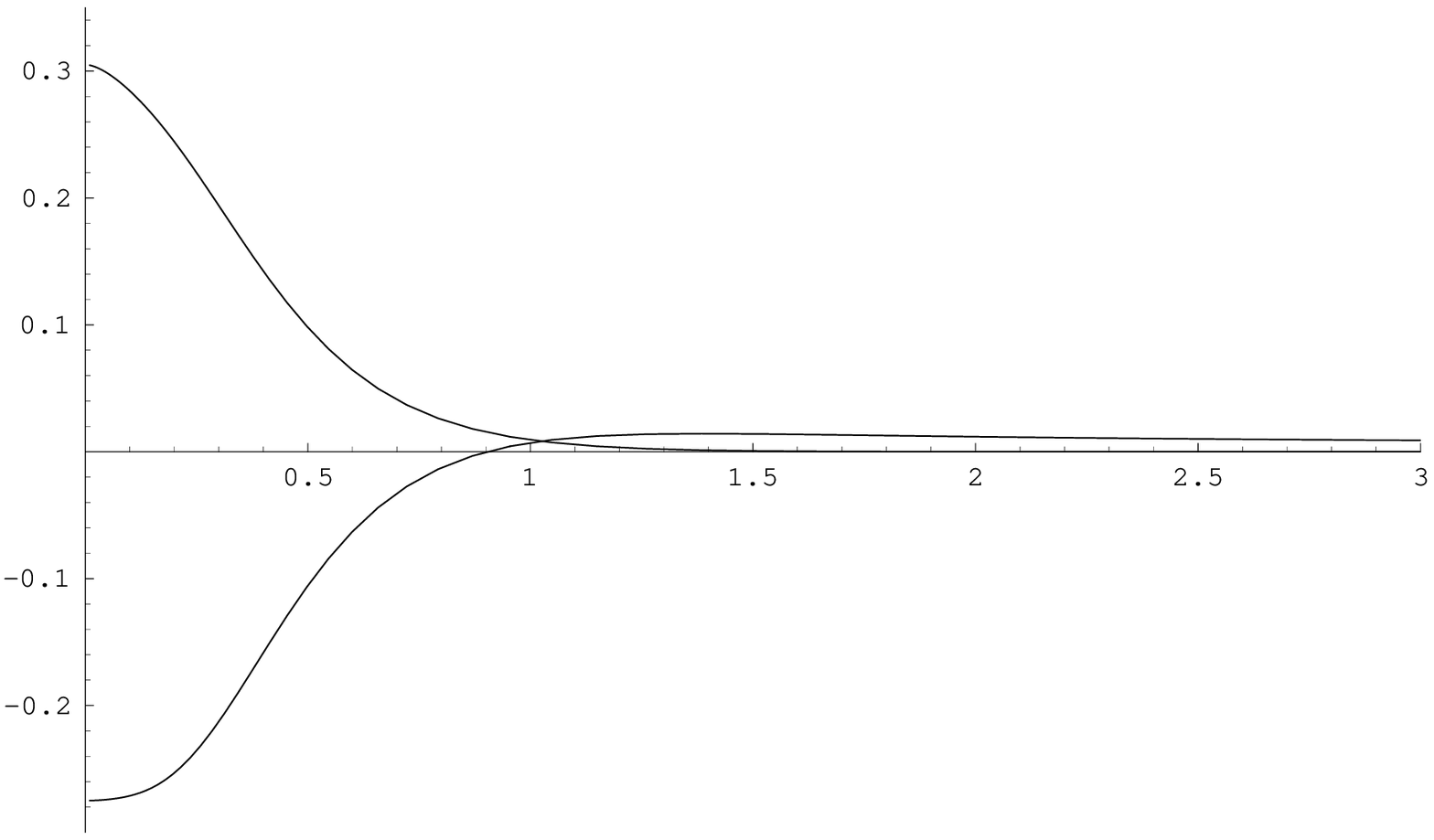,width=14cm,height=6cm}}
\put(1.5,5.0){${\cal E}^{spin}_{as}$}
\put(2,0.75){${\cal E}^{spin}_f$}
\put(14,2.7){$R$}
\end{picture}
\caption{ Spinor field. The curves of  the asymptotic and of the finite part of the energy multiplied by $R^{2}\cdot \beta^{-4}$, for $\beta =0.4$.} 
\end{figure}

\begin{figure}[ht]\unitlength1cm
\begin{picture}(6,6)
\put(-0.5,0){\epsfig{file=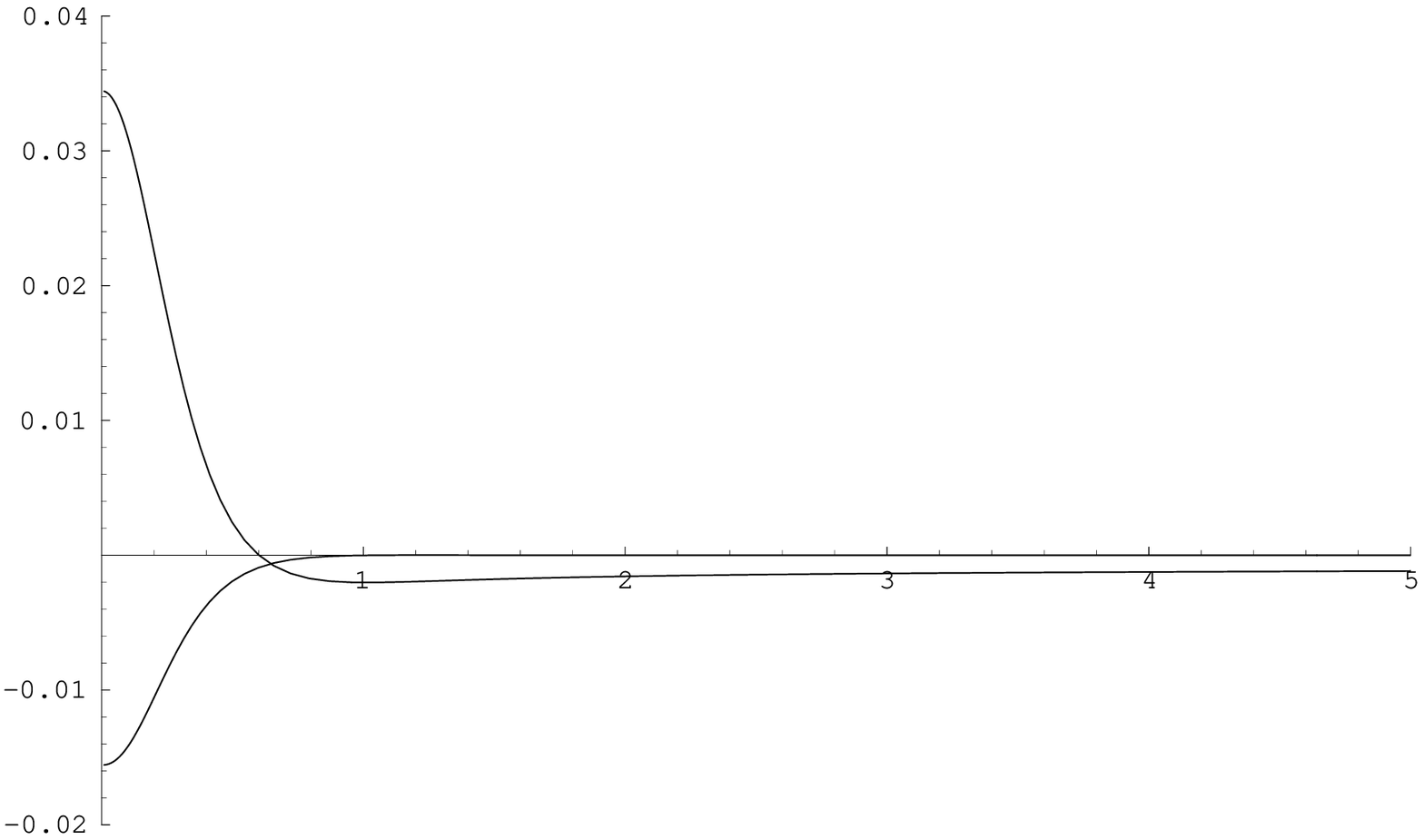,width=14cm,height=6cm}}
\put(1.1,5.0){${\cal E}^{spin}_{as}$}
\put(1.1,0.75){${\cal E}^{spin}_f$}
\put(14,2){$R$}
\end{picture}
\caption{Spinor field. The curves of  the asymptotic and of the finite part of the energy multiplied by $R^{2}\cdot \beta^{-4}$, for $\beta =2.2$.} 
\end{figure}

\begin{figure}[ht]\unitlength1cm
\begin{picture}(6,6)
\put(-0.5,0){\epsfig{file=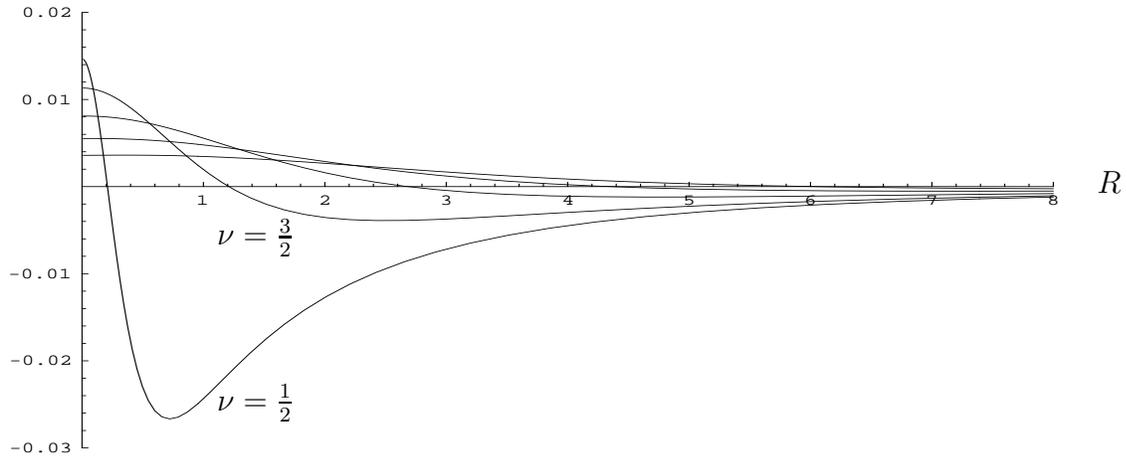,width=14cm,height=6cm}}
\put(2.3,2.8){$\nu=\frac 32$}
\put(2.3,0.60){$\nu=\frac 12$}
\put(14,3.5){$R$}
\end{picture}
\caption{Spinor field. The contributions $\nu=\frac 12,...\frac 92$ to the finite part of the energy multiplied by $R^{2}\cdot \beta^{-4}$, for $\beta =10$.} 
\end{figure}

\begin{figure}[ht]\unitlength1cm
\begin{picture}(6,6)
\put(-0.5,0){\epsfig{file=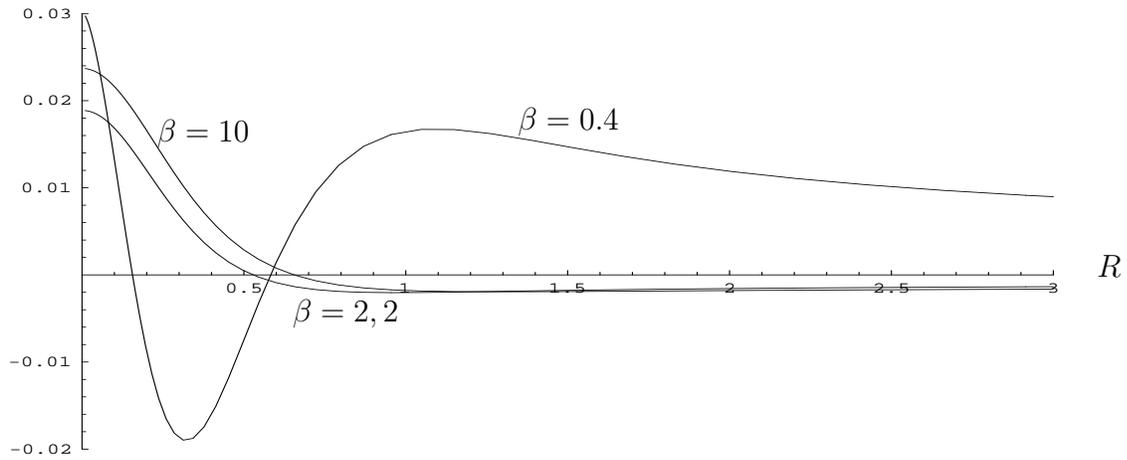,width=14cm,height=6cm}}
\put(1.5,4.2){$\beta=10$}
\put(6.3,4.35){$\beta=0.4$}
\put(3.3,1.8){$\beta=2,2$}
\put(14,2.4){$R$}
\end{picture}
\caption{Spinor field. The complete renormalized vacuum energy  ${\cal E}^{spin}(R)$ multiplied by $R^{2}\cdot \beta^{-4}$, for different values of strength of the 
potential.} 
\end{figure}

\end{document}